# Using Transcoding for Hidden Communication in IP Telephony

Wojciech Mazurczyk, Paweł Szaga, Krzysztof Szczypiorski
Warsaw University of Technology, Institute of Telecommunications
Warsaw, Poland, 00-665, Nowowiejska 15/19

**Abstract.** The paper presents a new steganographic method for IP telephony called TranSteg (Transcoding Steganography). Typically, in steganographic communication it is advised for covert data to be compressed in order to limit its size. In TranSteg it is the *overt data* that is compressed to make space for the steganogram. The main innovation of TranSteg is to, for a chosen voice stream, find a codec that will result in a similar voice quality but smaller voice payload size than the originally selected. Then, the voice stream is transcoded. At this step the original voice payload size is intentionally unaltered and the change of the codec is not indicated. Instead, after placing the transcoded voice payload, the remaining free space is filled with hidden data. TranSteg proof of concept implementation was designed and developed. The obtained experimental results are enclosed in this paper. They prove that the proposed method is feasible and offers a high steganographic bandwidth. TranSteg detection is difficult to perform when performing inspection in a single network localisation.

## 1. Introduction

Voice over IP (VoIP), or IP telephony, is one of the services of the IP world that is changing the entire telecommunication's landscape. It is a real-time service, which enables users to make phone calls through data networks that use an IP protocol. An IP telephony connection consists of two phases, in which certain types of traffic are exchanged between the calling parties (Fig. 1). These are:
- *Signalling phase* – in this phase signalling protocol messages, e.g. SIP messages (Session Initiation Protocol) [33], are exchanged between the caller and callee. These messages are intended to setup and negotiate the connection parameters between the calling parties.
- *Conversation phase* – if the previous phase is successful then the conversation takes place, in the form of two audio streams which are sent bidirectionally. RTP (Real-Time Transport Protocol) [34] is most often utilised for voice data transport, thus packets that carry voice payload are called RTP packets (see Fig. 2). The consecutive RTP packets form an RTP stream.

Steganography encompasses various information hiding techniques, whose aim is to embed a secret message (steganogram) into a carrier. Network steganography, to perform hidden communication, utilizes network protocols and/or relationships between them as the carrier for steganograms. Because of its popularity, IP telephony is becoming a natural target for network steganography [22]. Steganographic methods are aimed at hiding of the very existence of the communication, therefore any third-party observers should remain unaware of the presence of the steganographic exchange [35].

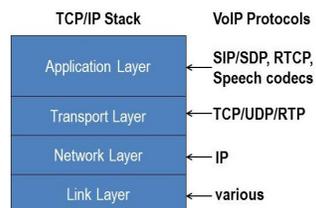

**Fig. 1:** Protocols for SIP-based VoIP

In this paper we introduce a new steganographic method – TranSteg (Transcoding Steganography) – which is intended for a broad class of multimedia and real-time applications, but its main foreseen application is IP telephony. TranSteg can also be exploited in other applications or services (like video streaming), wherever a possibility exists to efficiently compress (in a lossy or lossless manner) the overt data. The typical approach to steganography is to compress the *covert data* in order to limit its size (it is reasonable in the context of a limited steganographic bandwidth). In TranSteg compression of the *overt data* is used to make space for the steganogram – the concept is similar like in invertible authentication watermark that was first proposed by Fridrich et al. [10] for JPEG images. TranSteg bases on the general idea of transcoding (lossy compression) of the voice data from a higher bit rate codec (and thus greater voice payload size) to a lower bit rate codec (with smaller voice payload size) with the least possible degradation in voice quality.

The general idea behind TranSteg is as follows (the detailed procedures for different hidden communication scenarios are described in Sec. 3). RTP packets carrying user voice are inspected and the codec originally used for speech encoding (here called *overt codec*) is determined by analysing the PT (Payload Type) field in the RTP header (marked in Fig. 2).

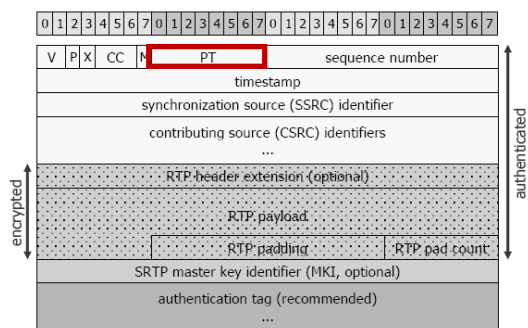

**Fig. 2:** RTP packet secured with SRTP protocol

Then, TranSteg finds an appropriate codec for the overt codec, called a *covert codec*. The application of the *covert codec* yields a comparable voice quality but a smaller voice payload size than originally. Next, the voice stream is transcoded, but the original, large, voice payload size and the codec type indicator are preserved, thus the PT field is left unchanged. Instead, after placing the transcoded voice of a smaller size inside the original payload field, the remaining free space is filled with hidden data (see Fig. 3).

It is worth noting that TranSteg can be utilized in different hidden communication scenarios – not only between the sender and the receiver of the RTP stream (i.e. in an end-to-end manner – see Sec. 3 for details). It can be also, in particular, utilized for secured VoIP streams e.g. using the most popular SRTP protocol [4] (Secure RTP) that provides confidentiality and authentication for RTP packets. Fig. 2 illustrates the parts of RTP packets that are encrypted and authenticated.

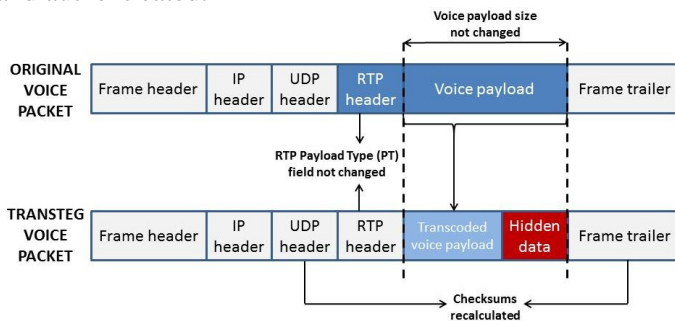

**Fig. 3:** Frame bearing voice payload: without (top) and with (bottom) hidden data inserted by TranSteg

TranSteg, like every steganographic method, can be described by the following set of characteristics: its steganographic bandwidth, its undetectability and the steganographic cost. The term – *steganographic bandwidth* – refers to the amount of secret data can be sent per time unit, when using a particular method. *Undetectability* is defined as the inability to detect a steganogram within a certain carrier. The most popular way to detect a steganogram is to analyse statistical properties of the captured data and compare them with the typical values for that carrier. Lastly, the *steganographic cost* characterises the degradation of the carrier caused by the application of the steganographic method. In the case of TranSteg, this cost can be expressed by means of providing a measure of conversation quality degradation, induced by transcoding and the introduction of an additional delay.

The contributions of this paper are:
- A detailed presentation of a new VoIP steganographic method,

- An analysis of the properties of TranSteg with the aid of a proof of concept implementation: the steganographic bandwidth, undetectability and the steganographic cost,
- Proposition of potential approaches for TranSteg detection.

The rest of the paper is structured as follows. Section 2 describes related work in VoIP steganography. Section 3 describes in detail the functioning of TranSteg and its hidden communication scenarios. Section 4 discusses the proof of concept implementation and presents the experimental results. Finally, Section 5 concludes our work.

## 2. Related Work

IP telephony as a hidden data carrier can be considered a fairly recent discovery. The proposed steganographic methods stem from two distinctive research origins. Firstly, from the well-established image and audio file steganography [5] – these methods targeted the digital representation of voice as carrier for hidden data. The second sphere of influence are the so called covert channels, created in different network protocols [1], [29] (a good survey on covert channels, by Zander et al., can be found in [43]) – these solutions target specific VoIP protocol fields (e.g. signalling protocol – SIP, transport protocol – RTP or control protocol – RTCP) or their behaviour. Presently, steganographic methods that can be utilized in telecommunication networks are jointly described by the term *network steganography*, or, specifically, when applied to IP telephony, by the term *steganophony* [22].

The first VoIP steganographic methods that utilize the digital voice signal as a hidden data carrier were proposed by Dittmann et al. in 2005 [7]. Authors had evaluated the existing audio steganography techniques, with a special focus on the solutions which were suitable for IP telephony. This work was further extended and published in 2006 in [19]. In [14], an implementation of SteganRTP was described. This tool employed least significant bits (LSB) of the G.711 codec to carry steganograms. Wang and Wu, in [41], also suggested using the least significant bits of voice samples to carry secret communication but here, the bits of the steganogram were coded using a low rate voice codec, like Speex. In [36], Takahashi and Lee proposed a similar approach and presented its proof of concept implementation – Voice over VoIP ($Vo^2IP$), which can establish a hidden conversation by embedding compressed voice data into the regular, PCM-based, voice traffic. The authors had also considered other methods that can be utilized in VoIP steganography, like DSSS (Direct Sequence Spread Spectrum), FHSS (Frequency-Hopping Spread Spectrum) or Echo hiding. Aoki in [2] proposed a steganographic method based on the characteristics of PCMU (Pulse Code Modulation), in which the 0-th speech sample can be represented by two codes due to the overlap. Another LSB-based method was proposed by Tian et al. in [40]. Authors had incorporated the m-sequence technique to eliminate the correlation among secret messages to resist statistical detection. A similar approach, also LSB-based, relying on adaptive VoIP steganography was presented by the same authors in [39]; a proof of concept tool - StegTalk – was also developed. In [27] Miao and Huang presented an adaptive steganography scheme that based on smoothness of the speech block. Such an approach proved to give better results in terms of voice quality than the LSB-based method.

Utilisation of the VoIP-specific protocols as a steganogram carrier was first proposed by Mazurczyk and Kotulski in 2006 [23]. The authors proposed using covert channels and watermarking to embed control information (expressed as different parameters) into VoIP streams. The unused bits in the header fields of IP, UDP and RTP protocols were utilized to carry the type of parameter and the actual parameter value is embedded as watermark into the voice data. The parameters are used to bound control information, including data authentication to the current VoIP data flow. In [25] and [26] Mazurczyk and Szczypiorski described network steganography methods that can be applied to VoIP: to its signalling protocol – SIP (with SDP), and to its RTP streams (also with RTCP). They discovered that a combination of information hiding solutions provides a capacity to covertly transfer about 2000 bits during the signalling phase of a connection and about 2.5 kbit/s during the conversation phase. In [26], a novel method called LACK (Lost Audio Packets Steganography) was introduced; it was later described and analysed in [24] and [22]. LACK relies on the modification of both: the content of the RTP packets, and their time dependencies. This method takes advantage of the fact that, in typical multimedia communication protocols, like RTP, excessively delayed packets are not used for the reconstruction of the transmitted data at the receiver, i.e. the packets are considered useless and discarded. Thus, hidden communication is possible by introducing intentional delays to selected RTP packets and substituting the original payload with a steganogram.

Bai et al. in [3] proposed a covert channel based on the jitter field of the RTCP header. This is performed two-stage: firstly, statistics of the value of the jitter field in the current network are calculated. Then, the secret message is modulated into the jitter field according to the previously calculated parameters. The utilization of such modulation guarantees that the characteristic of the covert channel is similar to that of the overt one. In [6], Forbes proposed a new RTP-based steganographic method that modifies the timestamp value of the RTP header to send steganograms. The method's theoretical maximum steganographic bandwidth is 350 bit/s.

The TranSteg technique presented in this paper is a development of the latter of the discussed groups of steganographic methods for VoIP, originating from covert channels. Compared to the existing solutions, its main advantages are a high steganographic bandwidth, low steganographic cost (i.e. little voice quality degradation) and difficult detection. However the last depends on the utilized hidden communication scenario. All of these features will be described and analysed in the context of TranSteg in the consecutive sections.

## 3. Communication Scenarios, Functioning and Detection

The performance of TranSteg depends, most notably, on the characteristics of the pair of codecs (as mentioned in the Introduction): one used originally to encode user speech – the *overt codec*, and one utilized for transcoding – the *covert codec*. It is worth noting that, depending on the hidden communication scenario, TranSteg may or may not be able to influence the choice of this codec. It is assumed that it is always possible to find a covert codec for a given overt one. However, it must be noted, that for very low bit rate codecs, the steganographic bandwidth shall be limited. In the ideal conditions the covert codec should:
- not degrade considerably user voice quality (caused by the transcoding operation and the introduced delays), when compared to the quality of the overt codec.
- provide the smallest achievable voice payload size as it results in the most free space in an RTP packet to convey a steganogram.

If there is a possibility to influence the overt codec (see the hidden communication scenarios below), in an ideal situation it should:
- result in a largest possible voice payload size to provide, together with the covert codec, the highest achievable steganographic bandwidth,
- be commonly used to not to raise suspicion.

Taking the above into account, TranSteg's steganographic bandwidth (*SB*) can be expressed as:

$$SB = (PS_O - PS_C) \cdot PN_S \quad [bit/s] \qquad (3\text{-}1)$$

where $PS_O$ denotes the overt codec's payload size, $PS_C$ is the covert codec's payload size and $PN_S$ describes the number of RTP packets sent during one second.

TranSteg can be utilized in four hidden communication scenarios (Fig. 4). The first scenario (S1 in Fig. 4) is the most common and typically the most desired: the sender and the receiver conduct a VoIP conversation while simultaneously exchanging steganograms (end-to-end). The conversation path is identical with the hidden data path. In the next three scenarios (marked S2-S4 in Fig. 4) only a part of the VoIP end-to-end path is used for hidden communication. As a result of actions undertaken by intermediate nodes, the sender and/or the receiver are, in principle, unaware of the steganographic data exchange. The application of TranSteg in IP telephony connections offers a chance to preserve users' conversation and simultaneously transfer steganograms. As noted previously, this is especially important for scenarios S2-S4.

In the abovementioned scenarios it is assumed that potential detection (steganalysis), usually executed by a warden [9], is not able to audit the speech carried in RTP packets because of the privacy issues related with this matter. Thus, the presence of a steganogram inside RTP packet payload can remain undiscovered. Other possibilities of TranSteg detection will be discussed in detail in subsection 3.3.

In the following part of this section TranSteg will be described with reference to the abovementioned scenarios. The most important factor in this context is whether the Steganogram Sender (SS) is located on the same host as the RTP packets' issuer. Thus, it may be able to control the RTP stream transmitter. Otherwise, when located on some intermediate network node, it will not be capable of such control.

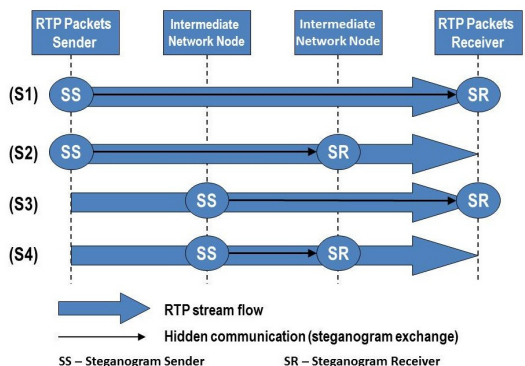

**Fig. 4:** Hidden communication scenarios for TranSteg

TranSteg may be also influenced by the utilization of SRTP protocol, which is used to provide the RTP stream with confidentiality and authentication. As mentioned in the Introduction, securing of the RTP stream does not necessarily impede the possibility of the exploitation of TranSteg. Such mode of operation may potentially even increase TranSteg's undetectability – this effect will be further investigated throughout the following subsections.

**3.1 Steganogram Sender controlling an RTP packet transmitter (scenarios S1 & S2)**

In scenario S1 a steganogram is embedded into an RTP packet and travels along the entire path between the RTP stream sender and receiver. Thus, there is no need to execute the operation of transcoding. User voice can be directly encoded with the desired covert codec with the omission of the prior encoding with the overt codec and thus avoid the whole process of transcoding. Despite this, the RTP stream will appear to have been encoded with the aid of the overt codec. The voice payload size and PT field in the RTP header shall not be changed. It is assumed that the SS and SR had agreed prior on the choice of the covert codecs corresponding to different overt codecs. Such common mapping may, for example, bind an overt codec G.711 with the covert codec G.726, or Speex 24.6 kbit/s with Speex 8 kbit/s, etc..

Thus, the SS shall perform the following steps for the embedding of a steganogram (Fig. 5):
- **Step 1:** Set the RTP payload size and modify Payload Type in the RTP header according to the chosen overt codec. These changes will indicate usage of the overt coding algorithm that will not, actually, be utilized.
- **Step 2:** The voice transcoded with the covert codec is inserted into the overt codec's RTP payload field.
- **Step 3:** Remaining free space is allocated for the hidden data and filled with a steganogram.
- **Step 4:** RTP packet is sent to the receiver.

When the modified TranSteg RTP stream reaches the SR, it extracts the voice payload and steganogram from the consecutive packets. The voice payload is then used for speech reconstruction and the steganogram parts are concatenated. This preserves the conversation functionality between the SS and SR and simultaneously enables hidden communication. For a third party observer, even if he/she is able to physically monitor the activity of both users (e.g. wiretap both locations) it will look like a regular call taking place.

To further mask the presence of TranSteg, SS can utilize the SRTP protocol to perform RTP payload encryption of both: the voice coded with a covert codec and the steganogram; thus making the detection of steganography even harder to perform (see Fig. 2).

The hidden communication scenario S1 offers most flexibility, and is advantageous when compared with the remaining ones, because:
- SS can choose the overt codec and thus influence the resulting steganographic bandwidth.
- The delays introduced by TranSteg to the RTP stream are the smallest in this scenario as there is no time consumption related with the transcoding (the voice is directly encoded with the covert codec).
- This scenario does not assume any required path of communication that the RTP stream should follow.

- To capacitate the exploitation of TranSteg, it is only necessary to modify the IP telephony client. Notably, the RTP protocol is usually implemented in software, which means it can be easily modified. No other protocol's modifications are required (i.e. UDP and frame checksums).
- RTP stream can be, additionally, secured with the aid of the SRTP protocol – this can be utilized to mask the contents of the transcoded voice data and the steganogram.

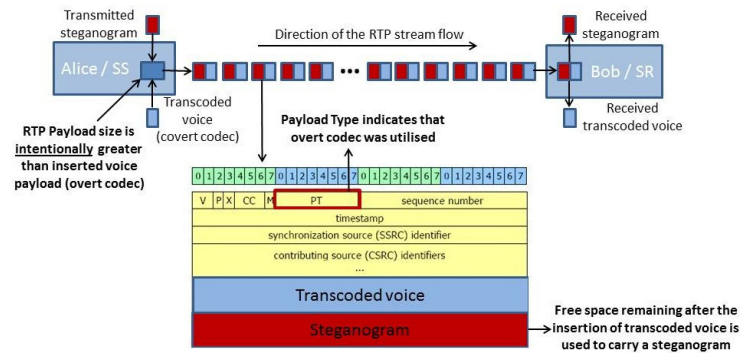

**Fig. 5:** The TranSteg concept – scenario S1 (SS – Steganogram Sender; SR – Steganogram Receiver)

In scenario S2, the main difference when compared with S1, is that the SR is situated at some intermediate network node. Thus, the IP telephony conversation is performed between the SS (caller) and an unaware of the steganographic procedure callee. The assumption in this scenario is that the SR is able to intercept and analyze all RTP packets exchanged between the SS and the callee. The TranSteg procedure for SS remains the same as in scenario S1. What changes is the behaviour of the SR.

When the tampered RTP stream reaches the SR, it performs the following steps:
- **Step 1:** It extracts voice payload and the steganogram from the RTP packets.
- **Step 2:** The voice payload is transcoded from the covert to overt codec and placed once again in consecutive RTP packets. By performing this task the steganogram is overwritten with user voice data.
- **Step 3:** Checksums for the lower layer protocols (i.e. the UDP checksum and CRC at the data link if they had been utilized) are adjusted.
- **Step 4:** Modified frames with encapsulated RTP packets are sent to the receiver (callee).

If the IP telephony connection is required to be secured with the SRTP protocol it does not impede the possibility to utilize TranSteg. The session keys used for authentication and encryption are exchanged between the calling parties before the conversation phase of the call and will be known in advance to the SS. This means that when the SS initiates an RTP stream, the first RTP packets contain transcoded voice but are intentionally *not* encrypted. Instead of a steganogram, they carry cryptographic keys that where negotiated between the SS and callee. Upon their extraction, the SR is able to encrypt the transcoded voice payload prior to forwarding it to the RTP packets receiver. Thus, the receiving party will not be aware of the steganographic procedure. Secondly, the SR will be capable of performing bidirectional hidden communication.

To summarize scenario S2:
- SS can still choose the overt codec and thus influence the resulting steganographic bandwidth.
- The delays introduced by TranSteg to the RTP stream depend on one transcoding operation.
- There is an assumption that SR is on the communication path between the calling parties and is able to oversee the whole RTP stream.
- TranSteg requires certain protocol modifications in the SR: the RTP and other network protocols (e.g. the UDP or data link layer protocols).
- Utilization of SRTP between the calling parties is not an obstacle for TranSteg. Analogically to scenario S1, it can be viewed as means to further mask hidden communication.

**3.2 Steganogram Sender Located at an Intermediate Network Node (scenarios S3 & S4)**

In scenario S3, the assumption is that SS is able to intercept and analyse all RTP packets exchanged between SR and the callee. SS does not control the RTP packet's transmitter, thus it cannot pick a suitable overt codec.

However, SR is a legitimate (overt) receiver of the RTP stream. Thus it is able to influence the choice of overt codec by negotiating it during the signalling phase of the call, with the calling party remaining unaware of the steganographic procedure. The behaviour of the SS is similar to the behaviour of SR in scenario S2 (see Sec. 3.1). The only difference is that SS is responsible for the transcoding from the overt to covert codec and for embedding of the steganogram – the remaining steps are the same. Thus, SS behaves as follows:

- **Step 1:** For an incoming RTP stream it transcodes the user's voice data from the overt to covert codec.
- **Step 2:** Transcoded voice payload is placed once again in an RTP packet.
- **Step 3:** The remaining free space of the RTP payload field is filled with steganogram's bits (thus the original voice payload is erased).
- **Step 4:** Checksums in lower layer protocols (UDP checksum and CRC at the data link) are adjusted.
- **Step 5:** Modified frames with encapsulated RTP packets are sent to the receiver (SR).

SR's operation is solely limited to extraction and analysis of the voice payload and steganogram from consecutive RTP packets (it is the same behaviour as in scenario S1, see Sec. 3.1).

In the presence of SRTP, in this scenario, the use of the TranSteg is not compromised – the conditions and the solution (cryptographic key's sharing between the SR and SS by means of TranSteg) is the same as in scenario 2 (see Sec. 3.1).

To summarize, in scenario S3:

- SR is responsible of influencing the choice of the overt codec by negotiating it with the calling party (unaware of the steganographic procedure).
- The delays introduced to the RTP stream by TranSteg depend on one transcoding operation.
- There is an assumption that SS is on the communication path between the calling parties and is able to oversee the whole RTP stream.
- TranSteg requires certain protocol modifications in the SS: the RTP and other network protocols (e.g. the UDP or data link layer protocols).
- Utilization of SRTP between the calling parties is not an obstacle for TranSteg. Analogically to scenario S1, it can be viewed as means to further mask hidden communication.

In scenario S4 it is assumed that both: SS and SR, are able to intercept and analyze all RTP packets exchanged between the calling parties. Thus, SS and SR cannot at all influence the choice of the overt codec, because they are both located at some intermediate network node (Fig. 6). Due to this fact they are bound to rely on the codec chosen by the overt, non-steganographic, calling parties. This, in particular, can result in low steganographic bandwidth as the hidden communication parties must adjust the covert codec to the negotiated overt codec. The most significant advantage of this TranSteg scenario is its potential use of aggregated IP telephony traffic to transfer steganograms. If both SS and SR have access to more than one VoIP call then the achievable steganographic bandwidth can be significantly increased, which can compensate for the loss in steganographic bandwidth caused by the inability to influence the choice of the overt codec.

The behavior of SS and SR is similar – they both perform transcoding: SS from overt to covert, and SR from covert to overt codecs. Steganogram is exchanged only along the part of the communication path where RTP stream travels "inside" the network – it never reaches the endpoints. The steps of the TranSteg scenario for SS are exactly the same as in scenario S3 (see above) and SR follows the logic presented in scenario S2 (see Sec. 3.1). It is also worth noting that, in this scenario, the utilization of SRTP protocol for conversation security entirely incapacitates the usage of TranSteg.

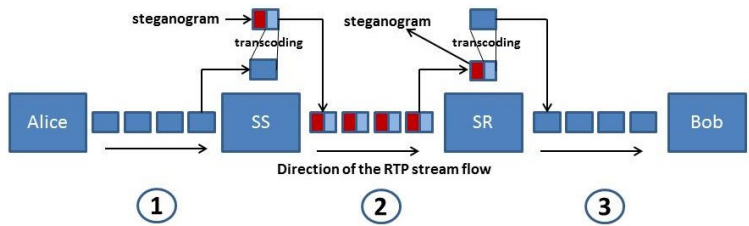

**Fig. 6:** The TranSteg concept – scenario S4 (SS – Steganogram Sender; SR – Steganogram Receiver)

To summarize scenario S4:
- There is an assumption that both: SS and SR, are on the communication path between the calling parties and are able to oversee the whole RTP stream.
- Neither SS nor SR can influence the choice of the overt codec, which potentially leads to a decrease in the steganographic bandwidth.
- There is a possibility to use aggregated VoIP traffic at the path between SS and SR, and thus significantly increase TranSteg's steganographic bandwidth.
- Neither SS nor SR are involved in the IP telephony conversation as overt calling parties. Thus, it is harder to detect hidden communication between the SS and SR comparing to the previously described scenarios (since neither is an initiator of the overt traffic).
- The delays introduced to the RTP stream are the highest compared with the other presented scenarios (due to the two transcoding operations).
- TranSteg requires certain protocol modifications in both: SS and SR; these involve modifications to the RTP and other network protocols (e.g. UDP or data link layer protocols).
- Utilization of SRTP to secure the conversation makes the use of TranSteg impossible.

**3.3 TranSteg Detection**

As mentioned at the beginning of Section 3, we assume that during the TranSteg-based hidden communication there is a warden executing detection (steganalysis) methods. We further assume that the warden will not be able to "listen" to the speech carried in RTP packets because of the privacy issues related with this matter. However, it must be emphasised that, if the SRTP protocol had been used for securing a TranSteg conversation, the warden will fail to detect the presence of steganograms in the RTP stream, in any of the below-mentioned scenarios (with the exception for S4).

To perform hidden communication, TranSteg utilizes modifications to the PDUs (Protocol Data Units) as a carrier – more precisely to the RTP payload field. When compared with other steganographic VoIP methods that, e.g. influence the order of the RTP packets or the delays between them, TranSteg does not introduce any irregularities to the RTP stream.

The successful detection of TranSteg mainly depends on:
- the location(s) at which the warden is able to monitor the modified RTP stream.
- the utilized TranSteg scenario (S1-4).

If the warden is capable of inspecting traffic solely in a single network, e.g. in the LAN (Local Area Network) of the overt transmitter or receiver, then the detection is hard to accomplish. The reason for the above is due to the fact that an RTP stream at a single traffic inspection point resembles legitimate streams. The remaining cases will be discussed below.

In scenario S1, there is no change of format of voice payloads during the traversing of the network. Thus, even if the warden would monitor traffic in different networks – the result would always be the same. Thus, the chances of TranSteg detection are very limited.

In scenarios S2 and S3 there is one transcoding operation, therefore the modification of the RTP packets' payload can be detected if the warden is able to probe and compare traffic from two localizations: prior and post the transcoding.

In scenario S4, there are three possible locations where the warden can inspect RTP traffic. They are marked as 1, 2 and 3 in Fig. 6. If a warden can monitor traffic in networks: 1 and 2 or 2 and 3 the detection capabilities are the same for scenarios S2 and S3. In the case when the warden is able to inspect the RTP stream in networks 1 and 3, where the voice format should be the same, then, due to the transcoding operation, some differences can be noted. This case is further investigated in Section 4. Thus, compared to the others, this scenario is the most susceptible to detection. Moreover, this scenario potentially can induce the largest voice quality degradation due to the necessary two transcoding operations.

Communication via TranSteg can be thwarted by certain actions undertaken by the wardens. The method can be defeated by applying random transcoding to every non-encrypted VoIP connection, to which the warden has access. Alternatively, only suspicious connections may be subject to transcoding. However, such an approach would lead to a deterioration of the quality of conversations. What must be emphasised, not only steganographic calls would be affected – the non-steganographic calls could also be "punished". In section 4 we provide guidelines for pinpointing suspicious IP telephony connections: we investigate RTP payload byte values' distribution as a possible indication of TranSteg utilization. It is worth noting that this approach will fail to succeed if SRTP protocol is applied.

Due to the above, it is necessary to explore other possibilities, which could facilitate the development of an efficient detection method fulfilling the constraints dictated by the VoIP's real-time operation constraints. One promising research direction worth pursuing is the adoption of the method proposed by Wright et al. in [42], which can be utilized for SRTP encrypted payload. However, this technique is only applicable for variable bit rate codecs. The authors of this work discovered that the lengths of encrypted RTP packets can be used to identify phrases spoken within a call. Therefore, if extended, this approach can be applied to deduce the characteristics of the employed speech codec, which would increase the probability of detection of covert communication.

The summary and comparison of hidden communication scenarios with respect to TranSteg functioning and detection (described in Sections 3.1-3.3) is presented in Table I.

**Table I** Comparison of hidden communication scenarios (S1-4)

| Scenario | S1 | S2 | S3 | S4 |
|---|---|---|---|---|
| SS/SR must be on the communication path to capture RTP stream | - | SR | SS | SS and SR |
| SS/SR influence the choice of the overt codec | SS and SR | SS | SR | - |
| Number of necessary TranSteg transcoding operations | 0 | 1 | 1 | 2 |
| Possibility to use aggregate VoIP traffic | - | - | - | + |
| Neither SS nor SR is a VoIP calling party (harder detection) | - | - | - | + |
| Necessary modifications to lower layer protocols, i.e. UDP checksum, frame CRC | - | + | + | + |
| SRTP utilization | Masks TranSteg | Masks TranSteg | Masks TranSteg | Prevents TranSteg |
| Number of necessary network monitoring/probing localizations for TranSteg detection | - | 2 | 2 | 2 |

## 4. TranSteg Implementation and Experimental Results

TranSteg implementation was developed for the hidden communication scenario presented in Fig. 6, i.e. when both SS and SR are located at intermediate network nodes. This scenario was chosen because, from the perspective of the delays introduced by TranSteg, it is a worst case scenario (due to the two transcoding operations - first at SS and then at SR). Thus, with the aid of the prototype, we want to find out to what extent TranSteg can degrade voice quality. For any other hidden communication scenario from Fig. 4, the introduced delays will be lower.

### 4.1 TranSteg Proof of Concept Implementation

TranSteg proof of concept implementation was developed in C++ on the Linux Ubuntu 10.10 (kernel 2.6.35) platform. Fig. 7 presents the functional architecture of the TranSteg prototype implementation. It is based on GTK+ 2.24.4 [12] threads which are used for GUI support, packet capture control and steganogram exchange. Netfilter [30] framework was also utilized for IP packet manipulation – modules: *iptables 1.4.10* and *libnetfilter_queue 0.0.17* were used.

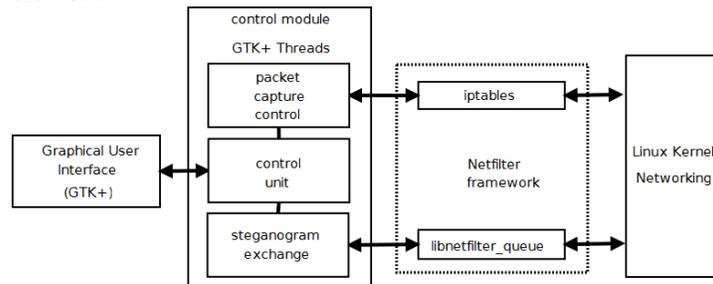

**Fig. 7:** TranSteg implementation architecture

All packets passing through a host are traversing *iptables* chains (see Fig. 8). There are three main types of these chains:

- INPUT – chain for incoming (received) packets that are intended for a process running on the local machine;
- OUTPUT – chain for packets that are being sent from a process running on the local machine;
- FORWARD – chain for packets that are being forwarded through the host (from one network interface to another).

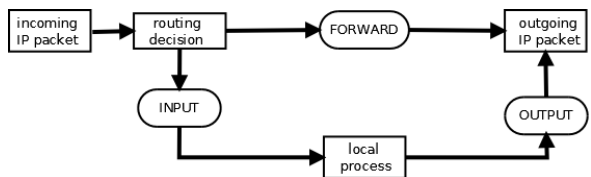

**Fig. 8:** IP packet traversing *iptables* chains

The developed TranSteg implementation enables controlling the *iptables* module. It is therefore possible to select from which chain the packets will be taken for modifications. This permits for the recreation of all of the previously discussed hidden communication scenarios (from S1 to S4) with the aid of the created prototype. Table II presents which chains should be used for each scenario S1-4.

**Table II** SS and SR *iptables* chains for different hidden communication scenarios S1-S4

| Scenario | SS *iptables* chain | SR *iptables* chain |
|---|---|---|
| S1 | OUTPUT | INPUT |
| S2 | OUTPUT | FORWARD |
| S3 | FORWARD | INPUT |
| S4 | FORWARD | FORWARD |

While traversing the *iptables* chains, packets are in the kernel space, which is inaccessible from the user space. Because of that, to perform modifications, packets from selected chains are put to a special QUEUE chain. This chain does not appear in the normal/usual packet traversing paths (see Fig. 8). Subsequently, the *libnetfilter_queue* module is used to send packets one by one to the user space, where TranSteg modifications take place. This process will be described further. Following the modifications, packets are returned to the QUEUE and continue traversing the *iptables* chains.

The TranSteg implementation bases on the G.711 (64 kbit/s) as an overt codec and G.726 (32 kbit/s) as the covert one. Thus, the transcoding is from G.711 to G.726 at the SS and the inverse is performed at the SR. The choice of the covert codec is entirely controlled by the SS and SR. Therefore, once the overt codec is known, a proper covert codec can be selected (e.g. one that does not degrade the voice quality and offers high steganographic bandwidth). In this TranSteg implementation G.711 codec was chosen as an overt codec because:

- It is the most popular speech codec that is widely used in IP telephony endpoints (both soft- and hard-phones) - it is simple to implement and does not require any license to be used.
- Generally, most common hard-phones and VoIP hardware devices, like gateways, (e.g. Cisco IP Phones, Sipura SPA-841, Vonage Phone Adapter, Linksys PAP2 or WRT54GP2) utilize codecs such as G.711, G.723 or G.729, while most of the popular softphones (e.g. SJPhone, Google Talk or X-lite) often offer, besides G.711, popular codecs like free Speex or iLBC (a list of VoIP clients and supported codecs can be found at: http://www.ozvoip.com/voip-codecs/devices/). In the majority of the endpoints G.711 is chosen as a default codec as it offers high quality – above 4 in MOS (Mean Opinion Score) scale – and is most likely to guarantee a successful speech codec negotiation during the signalling phase of a connection. For example, if we want to make a call between a Cisco IP Phone 7960 (which supports G.711 and G.729) and a popular, free softphone X-lite (which supports G.711, GSM, iLBC and Speex) successful connection negotiation is only possible with the G.711 codec.
- Whenever it is necessary to setup a call to PSTN (Public Switched Telephone Network) then, to avoid unnecessary transcoding and provide interoperability, the G.711 codec is frequently utilized.

### 4.2 Experiment Methodology

The experimental setup used to evaluate TranSteg's performance is presented in Fig. 9. The experimental environment was a controlled LAN network, so that no RTP packets were lost or excessively delayed. There we no network-related or endpoint-related interferences. Two hosts (A and B) took part in this experiment, both of them working under Linux Ubuntu 10.10 (kernel 2.6.35). On host A a soft-phone and TranSteg SS were launched, while on host B a soft-phone and TranSteg SR (Fig. 9).

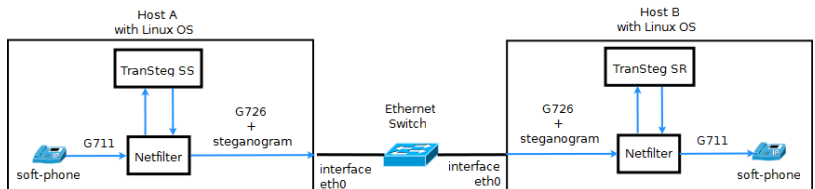

**Fig. 9:** TranSteg experimental test-bed

RTP packets were exchanged between the soft-phones (Ekiga 3.2.7 [8] and Linphone 3.3.2 [21] soft-phones were used in the tests). Both of them were configured to encode users' voice with the G.711 codec. After the generation of RTP packets at the originating user's soft-phone, they are taken by Netfilter module to the TranSteg SS application. Here they are being transcoded to G.726 and a steganogram is added. The transcoding is performed with the aid of Sun's CCITT implementation published on General Public License. RTP packet payload coded with G.711 is of the size 160 bytes. Post the transcoding to G.726, voice payload size decreases to 80 bytes. As mentioned earlier, TranSteg intentionally does not change the RTP payload field size after the transcoding. That is why the rest of the unused 80 bytes can be allocated for a steganogram. In the developed implementation, the steganogram is inserted into RTP packets from a user selected file. After changing the RTP payload, the application recalculates UDP checksum. Then the modified packet is returned to the Netfilter module and sent through outgoing network interface. Ethernet frame recalculation is done automatically before RTP packets are sent via the network interface. Then, after traversing the network, RTP packets reach host B. Next, they are redirected by the Netfilter module to TranSteg SR responsible for the extraction of the steganogram and transcoding the voice payload from G.726 back to G.711. It also inserts the transcoded voice into RTP payload fields and recalculates UDP checksums. Afterwards, the RTP packets are sent to soft-phone application for user conversation reconstruction.

Four TranSteg characteristics were measured with the aid of the above testbed (Fig. 10):
- Influence on the call quality (expressed in MOS scale),
- Steganographic bandwidth (expressed in kbit/s),
- Introduced delays (expressed in ms),
- Distribution of the byte values in RTP packets' payload.

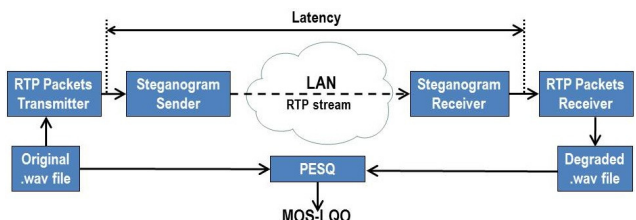

**Fig. 10:** TranSteg experimental setup

Call quality may be expressed in terms of subjective and objective quality measures. Objective measures are usually based on algorithms such as the E-Model [15], PAMS or PESQ [17] or others [20], [44]. In our analysis we shall use the subjective measure MOS (Mean Opinion Score) [16] calculated with the PESQ method. In our experiments, we used audio recordings from the TIMIT [11] continuous speech corpus - one of the most widely used corpora in the speech recognition community. Based on these recordings the voice packet payload was compiled into seven .wav files. Both male and female voices speaking English were used. Each resultant .wav input file was about 30 seconds long. The adopted coding involved PCM, 8000 Hz sampling, 16 bit sample representation of monophonic signal. It was ensured that the setup conformed to ITU-T P.862.3 recommendation [18] requirements which guarantees proper functioning of the PESQ method. Every .wav input file part was then inserted into the payloads of consecutive RTP packets and sent to the receiver where reassembling into a .wav file was performed. Then, the original and degraded files were compared with the use of PESQ and the resultant MOS-LQO (MOS-Listening Quality, Objective) was returned.

It was experimentally verified that the average call duration for IP telephony falls in the range of 7-11 minutes [13]. Thus, the tests for obtaining the values of steganographic bandwidth and latency were obtained for call duration of 9 minutes. The 9 minute representation was chosen to show how much secret data can be sent in one direction during a typical IP telephony call. Each experiment was repeated 20 times and the average results are presented.

Latency was measured with the aid of the MGEN 5.02 tool [37]. To achieve this goal MGEN utilizes NTP (Network Time Protocol) 4.2.6 [28], implementation v. 4.2.6 [31]. After the synchronization of two hosts: A and B (Fig. 11) latency measurements were performed.

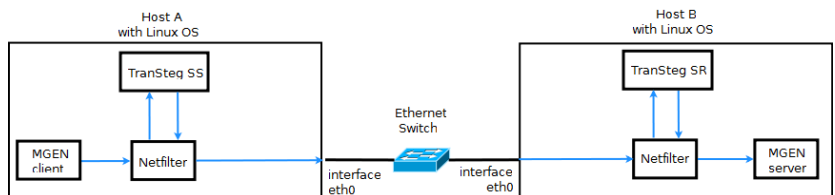
**Fig. 11:** Latency measurement

MGEN client and server were running on host A and B respectively. MGEN enables latency measurements by saving packet sending time in the packet payload. The MGEN client was sending a UDP stream, whose characteristic was chosen to exactly match RTP stream with G.711 payload (50 packets per second, 160 bytes of voice payload). The MGEN-generated packets were sent to the Netfilter module and then to TranSteg Steganogram Sender. In this module all of the TranSteg-related operations i.e. the transcoding of voice from G.711 do G.726 and the insertion of steganogram were performed, but packet payload was intentionally *not* changed. It was necessary to leave the original UDP payload because otherwise it would be impossible to extract packets' sending time. Next, packets were sent to host B. There they were transferred by the Netfilter module to the TranSteg Steganogram Receiver. In the SR all necessary operations i.e. transcoding from G.726 to G.711 and steganogram extraction were performed, and once again the payload was not changed. Then, when packets reached the MGEN server it would then generate a report about their sending and receiving time. Basing on the information from the MGEN report TRPR (TRace Plot Real-time) 2.1b2 tool [38] latency of each packet was calculated.

**4.3 Experimental Results**

The obtained experimental results are presented in Tables I, II and in Fig. 12-17. As mentioned in the previous subsection, TranSteg had been investigated for the hidden communication scenario S4 from Fig. 6.

Steganographic bandwidth for every performed call was identical and equalled 32 kbit/s. The explanation of this lies in the utilized codecs. For the overt codec – G.711 – the resulting payload is 160 bytes and for the covert codec – G.726 (32 kbit/s) – it is 80 bytes. Thus, when using 80 bytes for steganogram in every RTP packet (and 50 packets are generated every second of the call), then, during the whole, typical IP telephony connection it is possible to transfer about 2.2 MB (in each direction). This must be considered as a high steganographic bandwidth when compared with other existing VoIP steganographic methods.

The voice quality results (Table III) show that the average obtained result 3.83 (in MOS scale), is similar to the voice quality offered by G.726 (it is about 3.85). Thus, the resulting voice quality is lower than originally but it is still considered as good – the change is almost imperceptible for an average IP telephony user.

**Table III** TranSteg voice quality results

| Wav file number | 1 | 2 | 3 | 4 | 5 | 6 | 7 |
|---|---|---|---|---|---|---|---|
| Reference MOS | 4.46 | | | | | | |
| MOS-LQO | 4.013 | 3.908 | 3.687 | 3.657 | 3.709 | 3.715 | 4.149 |
| Average MOS-LQO | **3.834** (with standard deviation 0.18) | | | | | | |

Table IV presents delays introduced by TranSteg. The measured latency difference between calls with and without TranSteg turned out to be about 0.4 ms. This signifies that TranSteg does not introduce significant delays that could seriously affect voice quality. The exemplary latency results for VoIP calls with and without TranSteg are presented in Fig. 12. It is worth noting that latency was measured for the worst case scenario (from the point of view of the introduced delays) where two transcoding operations took place (first at SS, and then the second at SR). For the other hidden communication scenarios from Fig. 4 the introduced delays would be even lower. However, it must be emphasised that for a different pair of codecs the results would differ as both codecs: G.711 and G.726, are of low computational complexity.

**Table IV** TranSteg latency results

|  | **With TranSteg** | **Without TranSteg** | **Difference** |
|---|---|---|---|
| Average Latency [ms] | 1.24 | 0.85 | 0.39 |
| Standard Deviation [ms] | 0.32 | 0.07 | - |

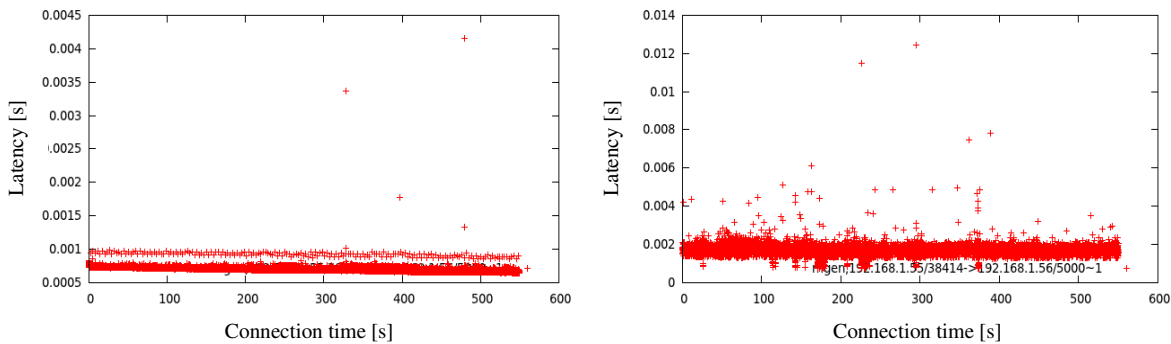

**Fig. 12:** Latency results for one exemplary IP telephony connection without (left) and with (right) TranSteg

Distribution of the byte values in RTP packet's payload was also investigated. This was done to verify how much the transcoding operations and the addition of the steganogram change the voice payload. This knowledge can be later utilized to aid the development of TranSteg detection method (see Sec. 3.3).

Fig. 13 presents the byte values' distribution for G.711 encoded speech before it reaches SS, i.e. prior to transcoding to G. 726, and after this operation (after leaving SS). As expected, there is a significant difference between the two presented curves. Thus, if the warden is able to monitor RTP traffic in the two networks (1 and 2 in Fig. 6), then the suspicious IP telephony connections can be discovered.

However, it must be emphasised, that in the other hidden communication scenarios from Fig. 4 (S1-3), when the SRTP protocol is utilized for conversation security, the byte frequency distribution in RTP payloads before and after transcoding will be similar (due to SRTP encryption) and, thus, TranSteg will be difficult to detect.

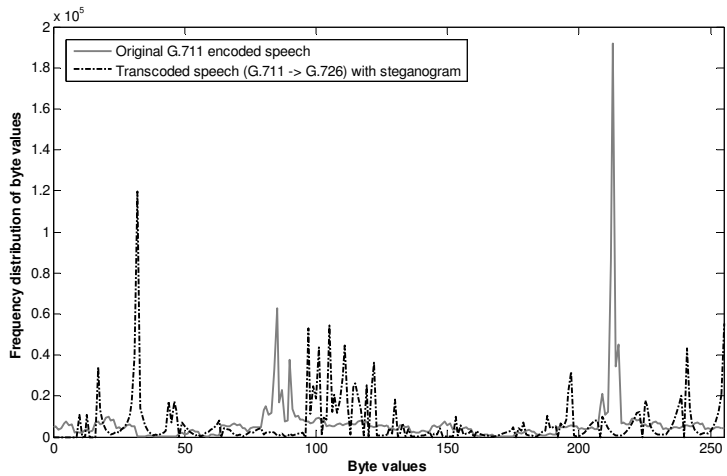

**Fig. 13:** Frequency distribution of byte values for G.711 encoded speech (before reaching SS) and transcoded to G.726 (after leaving SS)

Fig. 14 illustrates frequency distribution of byte values for the case where speech is transcoded at SS to G.726 – prior and post to the embedding of the steganogram. This diagram shows the sole influence of this operation, without the transcoding, of the voice payload. As can be seen, the two curve shapes are significantly different.

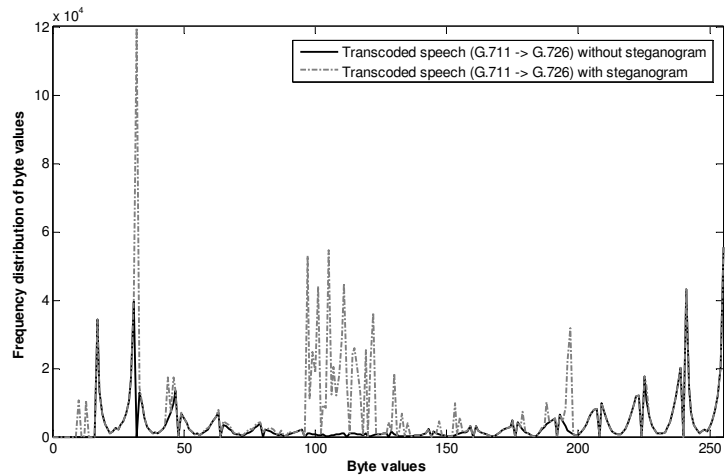

**Fig. 14:** Frequency distribution of byte values for speech transcoded to G.726 (at SS) without a steganogram and with a steganogram (after leaving SS)

In Fig. 15 it is presented how this situation can be changed when the same steganogram is compressed using zip algorithm. In this case, the two curve shapes are very similar. The compression operation leads to the randomization of the steganogram's bytes. This leads to an even shift in the observed frequency distribution of byte values.

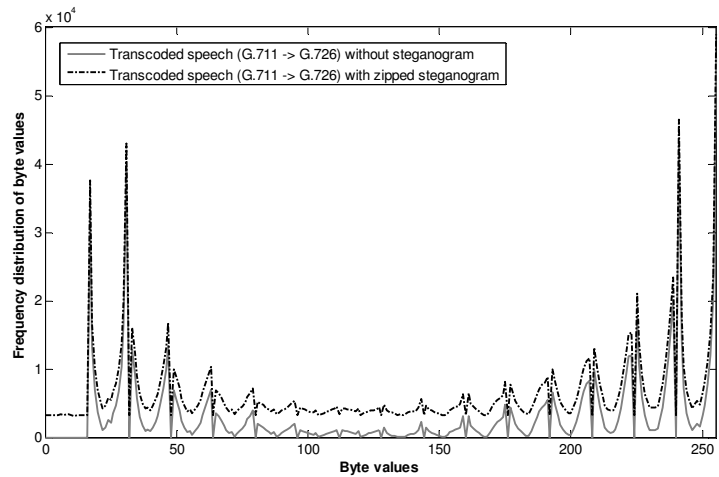

**Fig. 15:** Frequency distribution of byte values for speech transcoded to G.726 (at SS) without steganogram and with zipped steganogram (after leaving SS)

Finally, let us analyze for an originally G.711 encoded speech i.e. at a point prior to reaching SS and how frequency distribution of byte values changes past the inverse transcoding to G.711, i.e. after leaving the SR. The obtained results are illustrated in Fig. 16. Careful analysis of this figure leads to the conclusion that there are only slight differences between the G.711 encoded speech at the transmitter as compared with the received one (marked with arrows in Fig. 17). This means that, if the warden is able to monitor RTP traffic in two distinct localizations: in the overt transmitter's and the overt receiver's LANs (Local Area Networks), then TranSteg utilization could still remain unnoticed. The differences in shapes of the presented curves can be explained, to some extent, as introduced by the network, e.g. due to packet losses or transmission errors. The latter are present as the underlying transport protocol in the TCP/IP stack that is used is the unreliable UDP protocol. However, the presence of major differences can indicate TranSteg utilization.

The presented experimental results confirm the conclusions from Sec. 3.3. A warden capable of monitoring traffic in more than single network localization is very likely to detect the presence of TranSteg. However, it must be emphasized that the chosen hidden communication scenario S4 was selected to verify the potential influence of TranSteg on voice quality in a worst case scenario. From the point of view of the required effort for detection, this scenario is the easiest to defeat. It must be noted, that if any other scenario is utilized, e.g. the S1 scenario, together with SRTP encryption, then the disclosure of TranSteg can be very hard to attain.

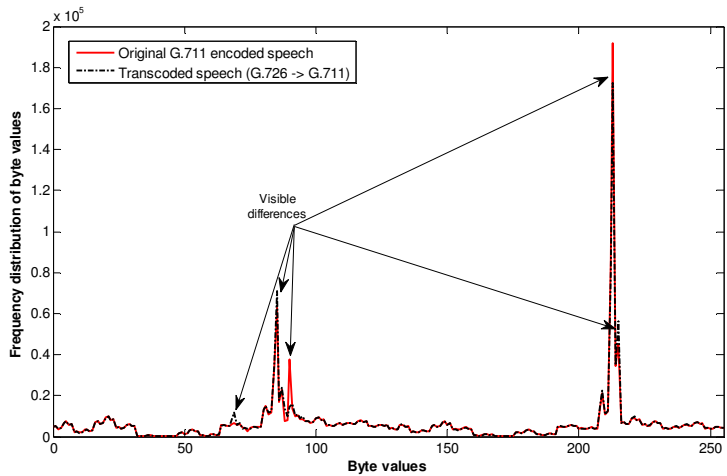

**Fig. 16:** Frequency distribution of byte values for G.711 encoded speech (before reaching SS) and transcoded to G.711 (after leaving SR)

However, even when considering the scenario S4, there is a simple way to obstruct the detection process, especially for the situation illustrated in Fig. 16. The proposed solution is to encode steganogram's bits until the original and transcoded speech's curves, at the receiver, are the same. Such an approach will surely limit the potential available steganographic bandwidth. At the same time, with the original and transcoded byte values' frequency distribution curves looking exactly the same, the disclosure of hidden communication shall be impossible.

To summarize, the detection of the TranSteg method is not trivial, especially for the hidden communication scenarios S1-3. Even for scenario S4 some simple measures can be taken to improve undetectability. Simple analysis of the frequency distribution of byte values in RTP payload as shown above may not be sufficient. Thus, as stated in Sec. 3.3, other possibilities leading to the development of an efficient detection method (one that would fulfil VoIP's real-time constraints) must be investigated.

## 5. Conclusions and Future Work

In this paper, a new IP telephony steganographic method, named TranSteg, was introduced. It was described basing on the possible hidden communication scenarios (S1-4 from Fig. 4). It was shown that the scenario where steganogram sender and receiver are the original source and final destination of the RTP traffic, respectively, is the most advantageous from the point of view of the achievable steganographic bandwidth, introduced steganographic cost and the undetectability of the method.

TranSteg proof of concept implementation was also designed and developed for a worst case scenario, where the introduced delays are largest. The obtained experimental results, for G.711 as an overt and G.726 as a covert codec, proved that the proposed method is feasible and offers a high steganographic bandwidth up to 32 kbit/s while introducing delays lower than 1 ms, and still retaining good voice quality (about 3.8 in MOS scale).

Detection of TranSteg strongly depends on the realized hidden communication scenario and the capabilities of a warden responsible for network steganography detection (e.g. the locations where it can monitor VoIP traffic). Generally, TranSteg detection can be difficult to perform, especially, if the SRTP protocol is utilized for securing RTP streams. Detection is also impeded when the warden is able to inspect traffic only in a single network localization.

Future work should involve an in depth analysis of speech codec pairs (overt and covert) that would be most advantageous for TranSteg. The algorithm for the selection of the covert codec will be developed, with the consideration of assuring an acceptable voice quality, low introduced delays and different VoIP codecs' characteristic features. Moreover, a prototype implementation should be developed for an end-to-end hidden communication scenario (S1 from Fig. 4) with the SRTP capability. This will allow the pursuing of an efficient, real-time, TranSteg detection method. On the other hand, to enhance the undetectability of TranSteg the different mechanisms of spreading the steganogram over the voice data instead of filling them in the end of the payload will be analysed in more detail. Additionally, detection methods that proved to be successful for digital images based on identifying double-compression [32] or compression artefacts should be considered and evaluated in TranSteg context. However, it must be noted that when for TranSteg scenario S1 (Fig. 4) is utilized together with SRTP voice stream encryption then utilization of such techniques for TranSteg detection will likely fail.

## ACKNOWLEDGMENTS

- This work was supported by the Polish Ministry of Science and Higher Education under grants: 504/M/1036/0248/2011 and IP2010 025470.
- The authors would like to thank Elżbieta Zielińska from Warsaw University of Technology (Poland) for helpful comments and remarks.